%% file: main.tex
\documentclass[a4paper,11pt]{article}
\usepackage{jinstpub} 
\usepackage{lineno}

\proceeding{TWEPP25 Topical Workshop on Electronics for Particle Physics\\
October 6th-10th, 2025\\
Rethymno, Crete, Greece}

\title{\boldmath Hardware-Accelerated GNN-based Hit Filtering for the Belle II Level-1 Trigger}

\author[a,1]{G. Heine,\note{Corresponding author.}}
\author[b]{F. Mayer,}
\author[b]{M. Neu,}
\author[b]{J. Becker,}
\author[a]{T. Ferber}
\affiliation[a]{Institute of Experimental Particle Physics, Karlsruhe Institute of Technology,\\Wolfgang-Gaede-Straße 1, 76131 Karlsruhe, Germany}
\affiliation[b]{Institute for Information Processing Technology, Karlsruhe Institute of Technology,\\Engesserstraße 5, 76131 Karlsruhe, Germany}

\emailAdd{greta.heine@kit.edu}

\abstract{\input{content/0_abstract}}


\input{acronyms}
\glsdisablehyper
\notoc

\begin{document}
\maketitle
\flushbottom
\glsresetall
\input{content/1_introduction}
\input{content/2_GNN}
\input{content/3_hardware}

\input{content/4_evaluation}
\input{content/5_conclusion}

\appendix
\acknowledgments
\input{content/acknowledgements}


\bibliographystyle{JHEP}
\bibliography{biblio.bib}



\end{document}

%% file: content/0_abstract.tex
We present a hardware-accelerated hit filtering system employing \glspl{gnn} on \glspl{fpga} for the Belle~II Level-1 Trigger. The \gls{gnn} exploits spatial and temporal relationships among sense wire hits and is optimized for high-throughput hardware operation via quantization, pruning, and static graph-building. 
Sector-wise spatial parallelization permits scaling to full-detector coverage, satisfying stringent latency and throughput requirements. 
At a sustained throughput of \SI{31.804}{\mega\hertz}, the system processes sense wire data in real-time and achieves detector-level background suppression with a measured latency of \SI{632.4}{\nano\second} while utilizing \SI{35.65}{\percent} of \glspl{lut}, and \SI{29.75}{\percent} of \glspl{ff}, with zero \gls{dsp} usage, as demonstrated in a prototype implementation for a single sector on an AMD Ultrascale XVCU190. Offline validation using Belle~II data yields a background hit rejection of \SI{83}{\percent} while maintaining \SI{95}{\percent} signal hit efficiency. This work establishes hit-level \gls{gnn}-based filtering on \glspl{fpga} as a scalable low-latency solution for real-time data reduction in high-luminosity collider conditions.

%% file: acronyms.tex
\newacronym{daq}{DAQ}{Data Acquisition}
\newacronym{l1}{TRG}{Level-1 Trigger}
\newacronym{hlt}{HLT}{High-Level Trigger}
\newacronym{fee}{FEE}{Front-End Electronics}
\newacronym{grl}{GRL}{Global Reconstruction Logic}
\newacronym{gdl}{GDL}{Global Decision Logic}
\newacronym{fpga}{FPGA}{Field-Programmable Gate Array}
\newacronym{cdc}{CDC}{Central Drift Chamber}
\newacronym{klm}{KLM}{K-Long and Muon detector}
\newacronym{ecl}{ECL}{Electromagnetic Calorimeter}
\newacronym{adc}{ADC}{Analog-to-Digital Converter}
\newacronym{tdc}{TDC}{Time-to-Digital Converter}

\newacronym{gnn}{GNN}{Graph Neural Network}
\newacronym{mva}{MVA}{Multi-Variate Analysis}
\newacronym{mlp}{MLP}{Multi-Layer Perceptron}
\newacronym{mc}{MC}{Monte Carlo}

\newacronym{lut}{LUT}{Look-Up Table}
\newacronym{ff}{FF}{Flip-Flop}
\newacronym{dsp}{DSP}{Digital Signal Processing}
\newacronym{pe}{PE}{Processing Element}
\newacronym{fifo}{FIFO}{First-In-First-Out buffers}

%% file: content/1_introduction.tex
\section{Introduction}
\label{sec:intro}
The Belle~II experiment \cite{abe2010belleiitechnicaldesign} at the SuperKEKB electron-positron collider (Japan) studies flavour and dark sector physics at the $\Upsilon$(4S) resonance energy of \SI{10.58}{\giga\electronvolt}. 
As SuperKEKB approaches an instantaneous luminosity of \SI{6e35}{\centi\metre^{-2}\second^{-1}}, the resulting rise in beam-induced backgrounds places increasing demands on the Belle~II trigger system.\\
The \gls{daq} system manages dataflow rates up to \SI{3}{\giga\byte\per\second} \cite{Yamada2015}, employing a two-level trigger system to process the approximately \SI{250}{\mega\hertz} bunch crossing rate.
The \gls{l1} system utilizes \gls{fpga} devices for real-time event selection with a total latency budget below \SI{4.4}{\micro\second} while reducing event rates to a maximum of \SI{30}{\kilo\hertz}.
The \gls{cdc} trigger~\cite{Taniguchi_2017} is one of four sub-trigger components that analyze data from dedicated sub-detector \gls{fee} boards and forward outputs to the global trigger system for the final trigger decision \cite{Lai_2025}. It reconstructs charged particle trajectories from \num{14336} sense wires per clock cycle of the $f_{sys}=\SI{127.216}{\mega\hertz}$ system clock, with an internal data clock of $f_{CDC}=\SI{31.804}{\mega\hertz}$, to derive kinematic track properties and suppress background.\\
Increasing luminosity directly raises detector occupancies and background levels. 
The average number of unassigned \gls{cdc} hits per event, $n_{\text{extraCDC}}$, has risen from about about 200 in 2021 to about 1900 in 2024, with projections up to about 2800 for future high-luminosity running.
The current Belle~II \gls{cdc} trigger algorithm mitigates these backgrounds by requiring aligned hits in a majority of the \gls{cdc} layers and down-scaling of the trigger rate of specific track topologies, measures that effectively suppress background but also constrain overall track finding efficiency.
Internal studies suggest even stricter criteria may be necessary in the future with rising background levels to stay within the \gls{daq} limits. \\
\Glspl{gnn} offer a promising solution for background mitigation by leveraging both individual hit characteristics and the spatial and temporal patterns in hit topologies. 
Recent work on deploying \glspl{gnn} for track reconstruction on \glspl{fpga} \cite{Dittmeier:2025nlh, loshchilov2019decoupledweightdecayregularization} and on ultra-fast architectures for jet tagging \cite{que2025jedilinearfastefficientgraph} has demonstrated the feasibility of real-time \gls{gnn} inference in high-rate collider environments.
While these works primarily focus on track-level reconstruction or jet tagging leveraging higher-level objects, our approach is distinguished by its focus on hit-level filtering directly on \gls{cdc} sense wire data. 
This enables real-time detector-level background suppression, achieving sub-microsecond latencies prior to track reconstruction in the Belle~II \gls{l1}. \\
This paper describes the \gls{gnn}-based hit filtering algorithm (\autoref{sec:gnn}) and its hardware implementation (\autoref{sec:hardware}), and evaluates performance in terms of hit classification efficiency, resource utilization, and latency (\autoref{sec:evaluation}), concluding in \autoref{sec:conclusion}.


%% file: content/2_GNN.tex
\section{GNN-based hit filtering}
\label{sec:gnn}

\begin{figure}[btp]
    \centering
    \begin{subfigure}{0.24\textwidth}
        \centering
        \includegraphics[width=\textwidth]{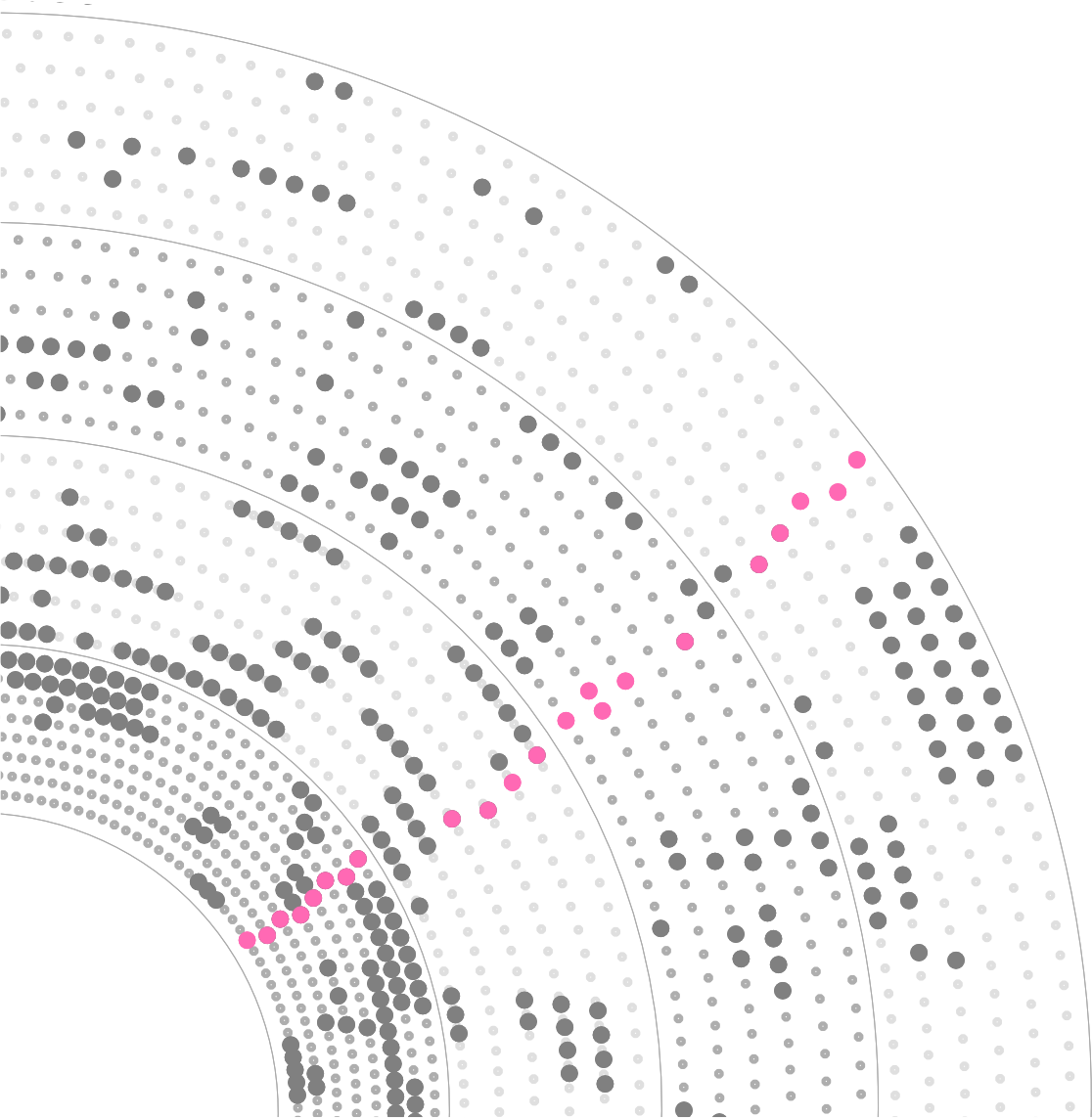} 
        \caption{before filtering}
        \label{fig:cleanup_steps_1}
    \end{subfigure}
    \hfill
    \begin{subfigure}{0.24\textwidth}
        \centering
        \includegraphics[width=\textwidth]{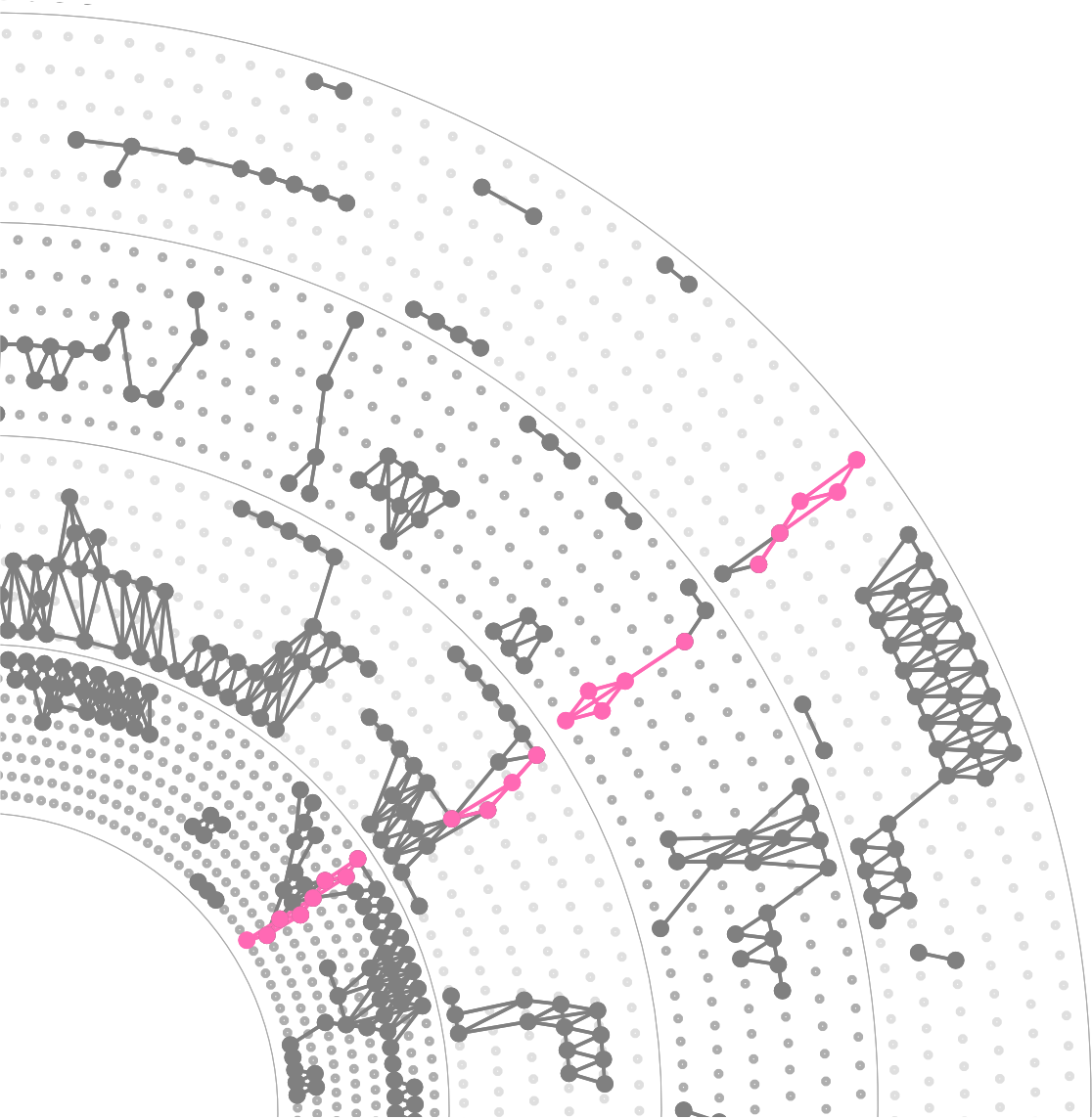} 
        \caption{graph representation}
        \label{fig:cleanup_steps_2}
    \end{subfigure}
    \hfill
    \begin{subfigure}{0.24\textwidth}
        \centering
        \includegraphics[width=\textwidth]{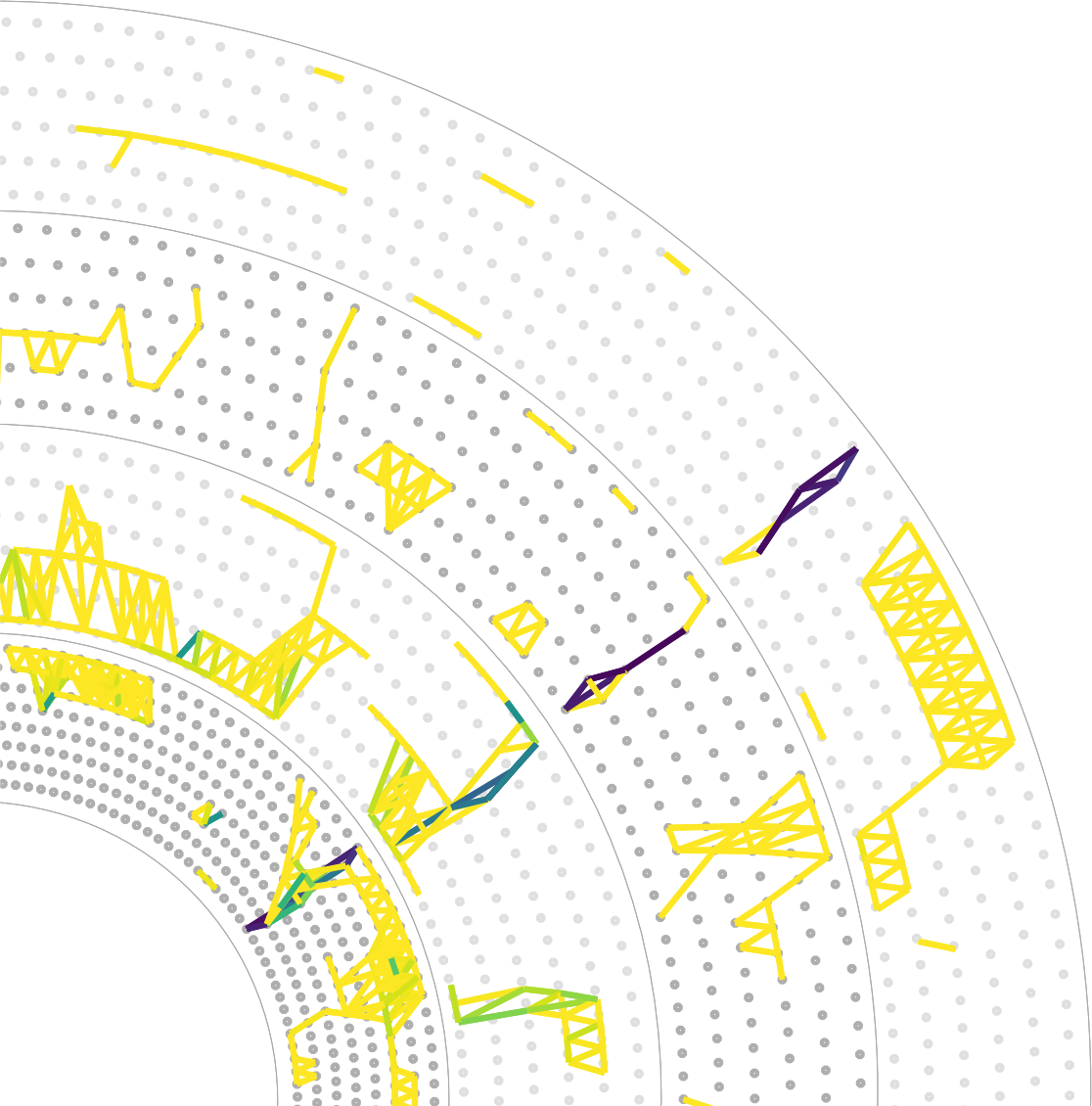} 
        \caption{GNN classification}
        \label{fig:cleanup_steps_3}
    \end{subfigure}
    \hfill
    \begin{subfigure}{0.24\textwidth}
        \centering
        \includegraphics[width=\textwidth]{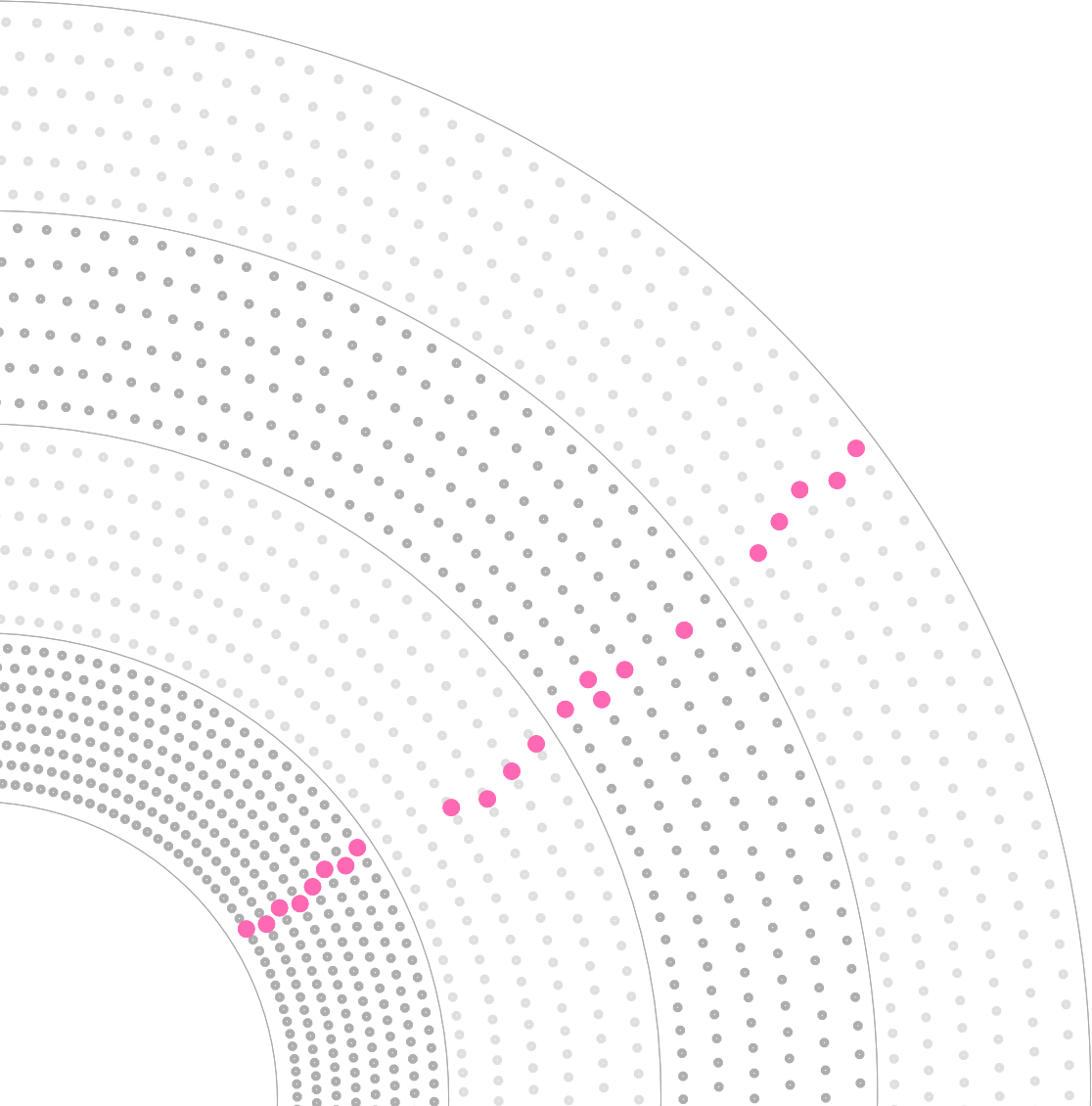} 
        \caption{after filtering}
        \label{fig:cleanup_steps_4}
    \end{subfigure}
    
    \caption{Overview of the \gls{gnn}-based hit filtering process: (\subref{fig:cleanup_steps_1}) \gls{cdc} hits before filtering with signal hits (pink) and background hits (grey), (\subref{fig:cleanup_steps_2}) hits are represented as graphs with edges connecting spatially compatible hits, (\subref{fig:cleanup_steps_3}) our \gls{gnn} performs edge or node classification to identify signal patterns where a dark colour denotes signal-like and light colour background-like classification, and (\subref{fig:cleanup_steps_4}) classification outputs are mapped back to individual hits for filtering.}
    \label{fig:cleanup_steps}
\end{figure}
We developed a \gls{gnn}-based hit filtering system developed for deployment in the Belle~II \gls{l1} system. The algorithm is designed with hardware constraints in mind but is first developed, trained, and validated in software using offline data. It involves a four-stage process shown in \autoref{fig:cleanup_steps} consisting of (1) reading and preprocessing of \gls{cdc} hits, (2) generating a graph representation of the hits, (3) applying \gls{gnn} classification on these graphs, and (4) a final hit filtering based on the \gls{gnn} classification scores. 

\subsection{Graph representation, network architecture and compression}
The \gls{cdc} hit data (\autoref{fig:cleanup_steps_1}) is transformed into a graph representation for \gls{gnn} processing (\autoref{fig:cleanup_steps_2}). 
Hits in each event become graph nodes, featuring $x$ and $y$ coordinates and \gls{adc} (sum of charge in the drift cell) values. Edges are formed using geometric rules to control connectivity and reduce computation \cite{Neu_2024}: 
(1) For same layer connections ($\Delta l=0$), wires with distance $\Delta w=\pm 1$ connect, (2) adjacent outgoing layers ($\Delta l = +1$) connect direct neighbours ($\Delta w = 0$), and (3) next-to-next layers ($\Delta l = +2$) connect wires offset by $\Delta w = -1, 0,$ or $+1$.
This pattern nearest-neighbour graph building mirrors the expected topology of particle trajectories. 
Edge features $\Delta r$, $\Delta \phi$, and $\Delta$TDC capture hit relationships, indicating if hits are originating from the same trajectory. 
Static graph features ($x$, $y$, $\Delta r$, $\Delta \phi$) are pre-computed in \gls{fpga} firmware to reduce online inference effort. \\
The \gls{gnn} architecture utilized in this work is a compressed variant of the Interaction Network framework \cite{battaglia2016interactionnetworkslearningobjects}. 
It comprises three \gls{mlp} blocks: two edge blocks $R1$ and $R2$, and a node block $O$, in addition to edge-to-node aggregation mechanisms. 
This architecture employs message-passing techniques for local pattern recognition within a two-node vicinity and is designed for minimal computational load.
The network can operate in both edge and node classification modes (\autoref{fig:cleanup_steps_3}), with node classification providing the best performance for the final hit filtering task (\autoref{fig:cleanup_steps_4}).\\
The subsequent compression methodologies transform the model into a variant that is optimized for hardware efficiency. 
By reducing the hidden layers' depth and width, the quantity of trainable parameters is decreased from 610 to 211. 
We implement quantization-aware training utilizing the Brevitas library \cite{brevitas}, employing 4-bit inputs, weights and activations, 16-bit biases, and 8-bit outputs. 
Pruning is executed to achieve \SI{50}{\percent} sparsity, which not only diminishes the network size but also unexpectedly enhances its performance. 
For \gls{fpga} deployment, the aggregation between $R1$ and $O$ is changed from sum to max to prevent overflow, and the sigmoid activation after $R2$ is replaced by a linear mapping to avoid costly computations. 
In total, these modifications reduce the network’s bit representation from \SI{19520}{bits} to \SI{797}{bits} per instance.

\subsection{Training strategy and dataset}
For training, we constructed a dataset inspired by \cite{Reuter_2025} to cover the full range of physics signatures and background conditions relevant to Belle~II operation. 
It consists of simulated \gls{mc} events with varied track topologies, overlaid with real beam-background data.
Additionally, real data samples derived from low-bias single-track triggered events~\cite{bähr2024neuralnetworkfirstlevelhardware} and \gls{hlt}-selected $\mu\mu(\gamma)$ events\footnote{\gls{hlt} $\mu\mu(\gamma)$ events are selected by requiring two oppositely charged, back-to-back tracks, each carrying momentum greater than $\SI{0.5}{\giga\electronvolt/c}$ in the center-of-mass reference frame and matched to an \gls{ecl} cluster with energy below $\SI{0.5}{\giga\electronvolt}$. The total energy of clusters, including possible photons, must be below \SI{2}{\giga\electronvolt}~\cite{The_Belle_II_Collaboration_Belle_II_Analysis}.
} are included. 
The ground truth labels for \gls{mc} samples utilize \gls{mc} truth information, whereas real data samples depend on offline track reconstruction to distinguish between signal and background classifications.
We apply a offline pre-filtering to the \gls{cdc} sense wire hits using their accumulated charge and timing digitized by the \gls{adc} and \gls{tdc} values, respectively, (\gls{adc}$\geq 10$, \gls{tdc} within \SI{500}{\nano\second} trigger time window) to maintain consistency with the established trigger pipeline and enhancing the signal-to-background ratio for more robust network training and inference.
Node and edge features are normalized to the range from -1 to +1 and undergo 4-bit quantization, matching trigger-level conditions.\\
The \gls{gnn} is implemented using PyTorch Geometric \cite{FeyLenssen2019}. 
To mitigate class imbalance, a binary cross-entropy loss function with an emphasis on positive sample weighting (weight=10) is employed.
The training runs for 50 epochs with a batch size of 1 and optimization facilitated by the AdamW \cite{loshchilov2019decoupledweightdecayregularization} optimizer. 
The utilized learning rate scheduler starts at 0.01 and decays by a factor of 0.7 every 4 epochs. A dropout rate of \SI{10}{\percent} is applied to reduce overfitting, and early stopping with a patience of 10 epochs terminates training once the validation performance plateaus.

%% file: content/3_hardware.tex
\section{Hardware implementation}
\begin{figure}[btp]
    \centering
    \includegraphics[width=1\linewidth]{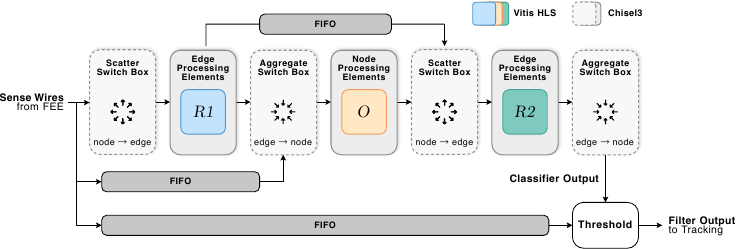}
    \caption{Block diagram of the hardware-accelerated Interaction Network architecture, where Vitis~HLS synthesized network blocks are mapped to dedicated \glspl{pe}. Static graphs generated from \gls{fee}-supplied sense wire data are propagated and updated via a series of scatter and aggregate Switch Boxes in-between \glspl{pe}, realized in Chisel into a register-transfer level design. The final classifier outputs, after threshold application, are sent to downstream tracking modules.}
    \label{fig:IN_hardware}
\end{figure}

As described in \autoref{sec:intro}, the \gls{l1} imposes strict latency and throughput requirements on the hit filtering algorithm. To fulfill these requirements, we deploy the previously shown Interaction Network as a dataflow accelerator on an \gls{fpga}. 
In order to facilitate the semi-automated implementation of the network on \gls{fpga}, we develop a deployment approach which transforms the compressed network into a register-transfer level design.
Our approach maps each layer of the neural network onto a dedicated \gls{pe}. 
In total, we design three types of \glspl{pe}: (1) Scatter Switch Boxes,  (2) Aggregate Switch Boxes, and (3) Network \glspl{pe}.
Switch Boxes are realized as hardware generators, implemented in the Chisel design language~\cite{bachrach:2012}. 
They embed the graph data structure into the dataflow accelerator, based on the approach described in \cite{Neu_2024}.
Network \glspl{pe} contain the trainable weights of the network, implemented using AMD~Vitis~HLS~\cite{vitis:2024:2}. To implement the \gls{mlp} blocks $R1$, $R2$ and $O$, we leverage architecture templates from the low latency HLS library~\cite{neu2025realtimegraphbasedpointcloud}.
Similar to hls4ml~\cite{fastml_hls4ml}, we define a reuse factor $R \in \{2^i : i \in \mathbb{N}^+\}$ which defines the spatial parallelism of the dataflow accelerator.\\
For our network described in \autoref{sec:gnn}, the resulting hardware architecture of our hit filtering algorithm is shown in \autoref{fig:IN_hardware}. 
\label{sec:hardware}
\noindent
\Glspl{pe} are shown in grey and are connected by either \glspl{fifo} or simple shift registers. 
All data interfaces are AXI4-Stream~\cite{ARM:IHI0051B} compliant and decoupled via ready-valid handshake.

%% file: content/4_evaluation.tex
\section{Performance evaluation}
\label{sec:evaluation}
\subsection{Hit filtering performance}
\label{subsec:filtering_performance}
We demonstrate the efficacy of the proposed \gls{gnn}-based hit clean-up through offline re-analysis of Belle~II data, focusing on \gls{hlt}-selected $\mu\mu(\gamma)$ events recorded at the end of 2024. 
As illustrated in \autoref{fig:result_hits}, we compare three configurations of the hit processing pipeline: the uncompressed full-precision \gls{gnn} model, its quantized 4-bit counterpart, and the unfiltered case. We run this simulation using sense wires of the full \gls{cdc} detector. 
For performance evaluation, hits are classified based on their association with offline reconstructed tracks. Signal hits are matched to tracks satisfying quality criteria\footnote{Track selection requires transverse momentum $p_\mathrm{T} > \SI{0.2}{\giga\electronvolt/c}$, total momentum $p > \SI{0.7}{\giga\electronvolt/c}$, longitudinal $|z_0| < \SI{15}{\centi\meter}$ and radial $|d_0| < \SI{15}{\centi\meter}$ distance from the interaction point, and at least 7 \gls{cdc} hits.}, while background hits $n_{\text{extraCDC}}$ comprise all unmatched hits, with no quality selection applied. Hits matched to tracks failing the quality criteria ($<\SI{5}{\percent}$ of signal, $<\SI{0.1}{\percent}$ of all hits), are excluded from performance metrics. Background rejection and signal efficiency are computed as the fraction of background and signal hits removed and retained, respectively.
Both the full-precision and 4-bit quantized models are applied at a working point corresponding to a \SI{95}{\percent} signal hit efficiency. 
The resulting distributions show an \SI{87}{\percent} background hit rejection for the full-precision model and \SI{83}{\percent} for the 4-bit version, demonstrating small degradation from model compression. 

\subsection{Experimental setup, resource utilization and timing}
\label{subsec:resources}
\begin{figure}[btp]
    \centering
    \begin{subfigure}{0.4570\textwidth}
        \centering
        \includegraphics[width=\textwidth]{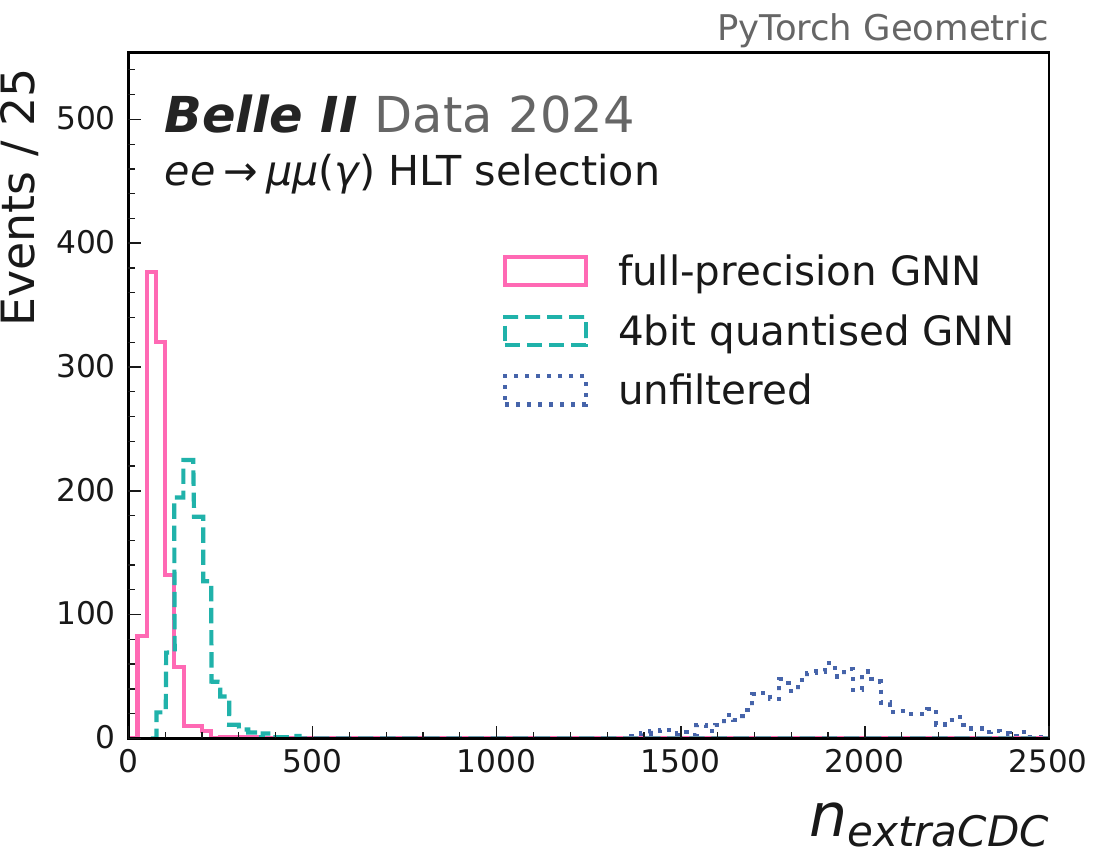} 
        \caption{Hit filtering performance}
        \label{fig:result_hits}
    \end{subfigure}
    \hfill
    \begin{subfigure}{0.5218\textwidth}
        \centering
        \includegraphics[width=\textwidth]{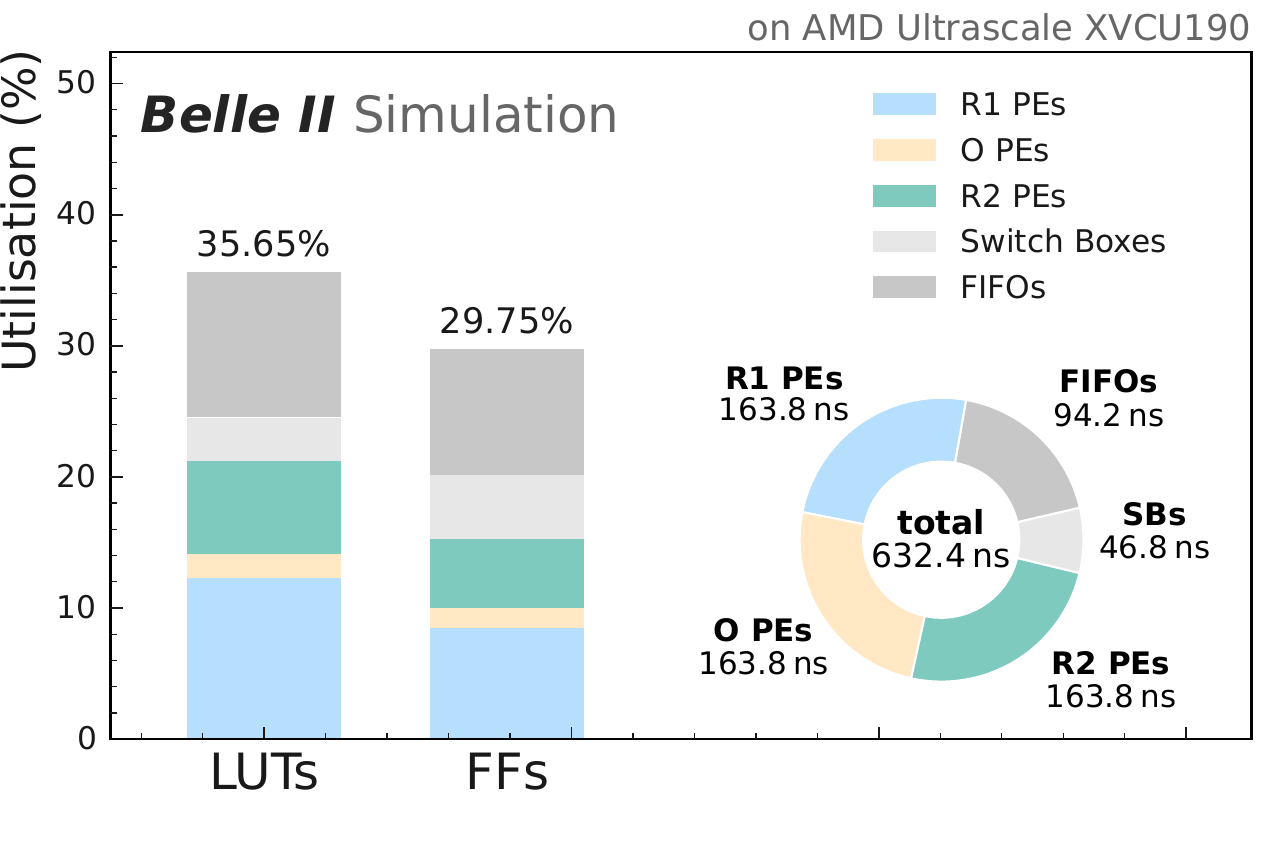}
        \caption{Resource utilization and latency}
        \label{fig:fpga_combined}
    \end{subfigure}
    \caption{\textbf{(\subref{fig:result_hits})} Background-like hit distributions $n_{\text{extraCDC}}$ processing Belle~II data (\gls{hlt}-selected $\mu\mu(\gamma)$ events from late 2024) with offline simulation on the full \gls{cdc} including and excluding the \gls{gnn} hit filtering. Both the full-precision and 4-bit quantised \gls{gnn} models achieve a background hit rejection of $>\SI{80}{\percent}$ at \SI{95}{\percent} signal hit efficiency. \textbf{(\subref{fig:fpga_combined})} \gls{fpga} resource utilization and latency per \gls{gnn} logic block for 495 sense wires and 2163 edges, showing modest \acrshort{lut}/\acrshort{ff} use and zero \glspl{dsp}. The total pipeline latency amounts to \SI{632.4}{\nano\second}; results are reported from Vivado~2024.2 after routing in out-of-context mode.}
    \label{fig:results}
\end{figure}

As a demonstrator for \gls{fpga} implementation, we use one of the 20 \gls{cdc} sectors described in \cite{Neu_2024}, comprising 495 sense wires and 2163 edges.
Defining the system frequency at $f_{GNN}=\SI{127.216}{\mega\hertz}$, in alignment with the \gls{l1} system clock, necessitates the processing of at least 124 nodes and 541 edges per clock cycle.
With a reuse factor $R = 4$, we achieve the required throughput of \SI{31.804}{\mega\hertz}.\\
We synthesize, place and route the register-transfer level design depicted in Figure~\ref{fig:IN_hardware} using AMD~Vivado~2024.2~\cite{vivado:2024:2} in out-of-context mode on an AMD Ultrascale XVCU190. 
For validation, we perform functional verification in a cycle-accurate simulation using CoCoTb~1.9.2~\cite{cocotb:2025} and ModelSim~2023.4~\cite{modelsim:2023:4}.  
The design meets all timing constraints with the processing latency and resource usage after routing detailed in \autoref{fig:fpga_combined}. 
The core accelerator components, including the edge block ($R1$, $R2$) and node block ($O$) \glspl{pe}, the Switch Boxes as well as \glspl{fifo} including buffers and queues show a total resource utilization of \SI{35.65}{\percent} for \glspl{lut} and \SI{29.75}{\percent} for \glspl{ff}, without using any \gls{dsp} blocks. 
The pipeline latency is \SI{163.8}{\nano\second} per major logic block, totalling \SI{632.4}{\nano\second} end-to-end.

 \noindent

%% file: content/5_conclusion.tex

\section{Conclusion}
\label{sec:conclusion}

We have developed a \gls{gnn}-based hit filtering system for the Belle~II Level-1 trigger.
Reprocessing 2024 Belle~II data with 4-bit quantized GNN implemented in PyTorch~Geometric shows \SI{83}{\percent} background hit rejection at \SI{95}{\percent} signal hit efficiency. 
We demonstrate the technical feasibility of our approach by deploying this network on an FPGA as a dataflow accelerator.
We verify functional correctness through cycle-accurate register-transfer level simulation and implement the design out-of-context on an AMD~Ultrascale~XCVU190.
Resource usage for a demonstrator sector is measured at \SI{35.65}{\percent} \glspl{lut}, \SI{29.75}{\percent} \glspl{ff}, and zero \glspl{dsp}, with a total pipeline latency of \SI{632.4}{\nano\second}, satisfying the Belle~II trigger constraints.
This implementation provides a viable basis for future integration into the Belle~II trigger system.


%% file: content/acknowledgements.tex
We thank the Belle II trigger team for their support and feedback. We also thank T.~Koga for comments on the manuscript, and G.~D.~Pietro for help with the Belle~II software framework.
This work is funded by BMFTR ErUM-Pro 05H24VK1.